\documentclass[12pt,a4paper]{article}
\usepackage[truedimen,margin=30mm]{geometry} 

\usepackage{mathrsfs}
\usepackage{amssymb}
\usepackage{amsmath}
\usepackage{ascmac}
\usepackage{amsthm}
\usepackage[dvips]{graphicx}
\usepackage{natbib}
\usepackage{setspace}
\usepackage{array}
\usepackage{times}
\usepackage{url}

\usepackage{color}
\usepackage{comment}

\usepackage{titlesec}
\titleformat*{\section}{\large\bfseries}
\titleformat*{\subsection}{\it}

\def\ep{{\varepsilon}}

\def\Nc{{\mathcal{N}}}
\def\E{\mathbb{E}}
\def\I{\mathbb{I}}

\title{
{\bf Sequential Adaptive Priors for \\
Orthogonal Functions}\footnote{\today}}

\date{}

\begin{document}

\maketitle
\doublespacing

\vspace{-1.5cm}
\begin{center}
{\large Shonosuke Sugasawa$^1$ and Daichi Mochihashi$^2$}

\medskip

\medskip
\noindent
$^1$Faculty of Economics, Keio University\\
$^2$The Institute of Statistical Mathematics
\end{center}

\vspace{0.5cm}
\begin{center}
{\bf \large Abstract}
\end{center}

We propose a novel class of prior distributions for sequences of orthogonal functions, which are frequently required in various statistical models such as functional principal component analysis (FPCA). Our approach constructs priors sequentially by imposing adaptive orthogonality constraints through a hierarchical formulation of conditionally normal distributions. The orthogonality is controlled via hyperparameters, allowing for flexible trade-offs between exactness and smoothness, which can be learned from the observed data. We illustrate the properties of the proposed prior and show that it leads to nearly orthogonal posterior estimates. The proposed prior is employed in Bayesian FPCA, providing more interpretable principal functions and efficient low-rank representations. Through simulation studies and analysis of human mobility data in Tokyo, we demonstrate the superior performance of our approach in inducing orthogonality and improving functional component estimation.


\section{Introduction}

Modeling a sequence of orthogonal functions is necessary for a variety of statistical problems. 
Examples include functional principal component analysis (FPCA) \citep{margaritella2021parameter,suarez2017bayesian}, interactive fixed effects for panel data \citep{bai2009panel} and Karhunen-Lo\`eve expansion for stochastic processes \citep{yao2005functional}. 
Such models entail a sequence of orthogonal functions to achieve interpretable and efficient representations,
especially for selecting an effective number of components from data.
Although a Bayesian framework needs to specify a prior distribution over such a sequence of functions, introducing priors that respect orthogonality constraints is nontrivial. 
For instance, existing Bayesian FPCA typically place priors without explicitly enforcing orthogonality \citep[e.g.,][]{suarez2017bayesian}. 
While such formulations offer computational convenience, they may compromise interpretability and efficiency of representation, which are central motivations for modeling with orthogonal functions in the first place.

There is a rich literature on incorporating constraints into Bayesian models, particularly in the context of shape-constrained function estimation. 
For example, monotonicity and convexity constraints have been widely studied in Bayesian nonparametric regression \citep{shively2009bayesian, lin2014bayesian,  lenk2017bayesian}, along with other structural constraints such as unimodality and bounded variation \citep{wheeler2017bayesian, yu2023bayesian}. 
While these works focus on constraints imposed on single functions, enforcing mutual orthogonality among multiple functions poses additional challenges due to the rapidly increasing number of pairwise constraints.
Several approaches have been proposed to encourage diversity among function components. 
\cite{plumlee2018orthogonal} proposed constraint Gaussian process to ensure orthogonality of a single function and pre-specified multiple functions.
Constraint relaxation methods \citep{duan2020bayesian, matuk2022bayesian} introduce soft penalties to approximate orthogonality, but these typically rely on a fixed tuning parameter, which may require careful calibration and cannot be adaptively learned from data. 
As related approaches, repulsive priors \citep{petralia2012repulsive, xie2020bayesian} can promote separation between functions but do not directly ensure orthogonality in the function space. 
Other alternatives include orthogonal random Fourier features \citep{yu2016orthogonal}, which produce random orthogonal basis functions but do not preserve orthogonality when used to construct multiple functional components. 
Similarly, models defined on the Stiefel manifold \citep[e.g.,][]{jauch2025prior} provide a principled way to impose orthogonal constraints through matrix-valued priors, though these can be computationally demanding and are often impractical when the basis functions themselves are not orthogonal. Post-hoc approaches that modify posterior samples to satisfy constraints \citep[e.g.,][]{sen2018constrained} offer a practical alternative, but still lack theoretical guarantees.

In this work, we propose the Adaptive Orthogonal (AO) prior, a novel prior distribution that directly induces approximate orthogonality among functions sequentially, in a computationally efficient and flexible manner. 
Our paper builds upon the framework of basis function expansions, where each function is represented as a linear combination of pre-specified basis functions. 
This formulation provides a flexible and interpretable structure that has been widely adopted in Bayesian modeling \citep[e.g.][]{lang2004bayesian} and offers a unified representation of functions, whose infinite-dimensional limit is known as Gaussian processes \citep{kimeldorf1970correspondence, williams2006gaussian}. 
Based on the basis function representation, our AO prior sequentially defines prior distribution for the coefficient vectors of each function using conditionally Gaussian distributions with approximate orthogonality constraints. 
By controlling the strength of these constraints through hyperparameters, our prior flexibly accommodates both exact and approximate orthogonality, allowing researchers to balance strictness and model complexity depending on the application.
AO prior can be seamlessly incorporated into Bayesian functional principal component analysis (FPCA), where orthogonality among principal component functions is crucial for both interpretability and dimensionality reduction. 
Our prior not only guarantees nearly orthogonal posterior estimates of the principal functions but also enables automatic selection of the effective number of components owing to the sequential construction of AO prior, leading to parsimonious low-rank representations. Furthermore, the conditional Gaussian formulation allows for straightforward implementation using standard Gibbs sampling, making posterior inference computationally tractable even in higher-dimensional settings.

This paper is organized as follows. 
In Section 2, we introduce the AO prior and discuss its theoretical properties. Section 3 presents Bayesian FPCA model incorporating the proposed prior. 
We demonstrate numerical performance of the proposed method through simulation studies in Section~4 and an application to human mobility data in Tokyo in Section~5. 
Finally, concluding remarks are given in Section 6.
The R code implementing the proposed method is available at GitHub repository (\url{https://github.com/sshonosuke/AOP}).

\section{Prior for Orthogonal Function Sequences}

\subsection{Prior construction}

We wish to construct priors for a set of orthogonal functions, $f_1(x),\ldots,f_K(x)$. 
To this end, we consider basis function expansions known as a flexible nonparametric approach for function estimation \citep[e.g.][]{lang2004bayesian}.
For example, Gaussian process models can be viewed as the infinite-dimensional limit of basis function models \citep{kimeldorf1970correspondence,williams2006gaussian}, where prior on functions is induced through a stochastic process equivalent to a (possibly infinite) linear combination of basis functions.
Let $\phi_l(x)$ be a basis function for $l=1,\ldots,L$, where $L$ is the number of basis functions. 
Define $\Phi(x)=(\phi_1(x),\ldots,\phi_L(x))$ as the $L$-dimensional vector of basis functions, and we model each function as $f_j(x)=\beta_j^\top \Phi(x)$, a linear combination of the pre-specified basis functions $\phi_l(x)$. 
Therefore, for $j=1,\ldots,K$ we want to construct a prior for $\beta_1,\ldots,\beta_K$ such that $K$ functions, $\beta_1^\top \Phi(x),\ldots,\beta_K^\top \Phi(x)$ become (nearly) orthogonal. 
For $j\neq k$, the inner product of $\beta_j^\top \Phi(x)$ and $\beta_k^\top \Phi(x)$ is given by 
$$
\int f_j(x)f_k(x)dx=\int \beta_j^\top \Phi(x) \beta_k^\top \Phi(x) dx=\beta_j^\top \Omega \beta_k,
$$
where $\Omega=\int \Phi(x)\Phi(x)^\top dx$.
Then, the condition that $f_j(x)$ and $f_k(x)$ are orthogonal is equivalent to 
\begin{equation}\label{eq:constraint}
 \beta_j^\top \Omega \beta_k=0 \,.
\end{equation}

We then sequentially define the joint prior of $(\beta_1,\ldots,\beta_K)$.  
First, the prior of $\beta_1$ is defined as  $\beta_1\sim \Nc_L(b_{01}, B_{01})$ without any restrictions, where $b_{01}$ and $B_{01}$ are hyperparameters and $\Nc_L(\mu, \Sigma)$ denotes the $L$-dimensional normal distribution with mean vector $\mu$ and variance-covariance matrix $\Sigma$.
Next, the prior of $\beta_2$ is defined, conditionally on $\beta_1$, as $\beta_1^\top \Omega \beta_2\sim \Nc(0, \tau_2^2)$ and  $H_2\beta_2\sim \Nc_{L-1}(b_{02}, B_{02})$, assuming prior independence between $\beta_1^\top \Omega \beta_2$ and $H_2\beta_2$, where $b_{02}$, $B_{02}$ and $\tau_2^2$ are hyperparameters and $H_2$ is $(L-1)\times L$ matrix with rank $L-1$ that is independent of $\beta_1$. 
Note that the constraint (\ref{eq:constraint}) exactly holds when $\tau_2=0$. 
This conditional prior specification is equivalent to assigning a multivariate normal prior as $A_2(\beta_1)\beta_2|\beta_1\sim \Nc(b_{02}^{\ast}, {\rm blockdiag}(\tau_2^2, B_{02}))$ where $b_{02}^{\ast}=(0, b_{02}^\top)^\top$ and $A_2(\beta_1)=((\beta_1^\top\Omega)^\top, H_2^{\top})^{\top}$.
Note that the first row vector of $A_2(\beta_1)$ is $\beta_1^\top\Omega$ and the other part is equivalent to $H_2$. 
Then, the conditional prior of $\beta_2$ given $\beta_1$ is given by
$$
\beta_2\mid \beta_1 \sim \Nc_L\Big(A_2(\beta_1)^{-1}b_{02}^{\ast}, A_2(\beta_1)^{-1}{\rm blockdiag}(\tau_2^2, B_{02})\{A_2(\beta_1)^\top\}^{-1}\Big),
$$
where $b_{02}^{\ast}=(0, b_{02}^\top)^\top$.
Similarly, given $\beta_{1:j}\equiv (\beta_1,\ldots,\beta_j)$, the conditional prior for $\beta_{j+1}$ is defined as $\beta_{k}^\top\Omega \beta_{j+1}\sim \Nc(0, \tau_{j+1}^2)$ independently for $k=1,\ldots,j$ and $H_{j+1}\beta_{j+1}\sim \Nc_{L-j}(b_{0,j+1}, B_{0,j+1})$, assuming prior independence among $\beta_{k}^\top\Omega \beta_{j+1}\ (k=1,\ldots,j)$ and $H_{j+1}\beta_{j+1}$, with hyperparameters $b_{0,j+1}$, $B_{0,j+1}$ and $\tau_{j+1}^2$.
Here $H_{j+1}$ is an $(L-j)\times L$ matrix that is independent of $\beta_{1:j}$.  
Again, this prior formulation is equivalent to assigning a multivariate normal prior for $A_{j+1}(\beta_{1:j})\beta_{j+1}$, where $A_{j+1}(\beta_{1:j})=((\beta_1^\top\Omega)^\top,\ldots,(\beta_j^\top\Omega)^\top, H_{j+1}^{\top})^{\top}$, that is, the first $k$th row vector is $\beta_k^\top \Omega$ for $k=1,\ldots,j$, and the other part is $H_{j+1}$.
Hence, the conditional prior of $\beta_{j+1}$ given $\beta_{1:j}$ is given by
\begin{equation}\label{eq:cond_prior}
\beta_{j+1}\mid \beta_{1:j} \sim \Nc_L\Big(A_{j+1}^{-1}b_{0,j+1}^{\ast}, A_{j+1}^{-1}{\rm blockdiag}(\tau_{j+1}^2 I_{j}, B_{0,j+1})(A_{j+1}^\top)^{-1}\Big),
\end{equation}
for $j=1,\ldots,K-1$, where $A_{j+1}\equiv A_{j+1}(\beta_{1:j})$ and $b_{0,j+1}^{\ast}=(0_j^\top, b_{0,j+1}^\top)^\top$.

The specification of $b_{0,j}$ and $B_{0,j}$ can be arbitrary. 
For example, if we set $b_{0,j}=0_{L-j}$ and $B_{0,j}=\gamma I_{L-j}$ for some large value of $\gamma$, leading to a diffuse prior.
On the other hand, $\tau_j^2$ controls the exactness of the orthogonal constraints; the constraints are exact when $\tau_j^2=0$.
In many applications, it is important to give nearly exact constraints on $f_1(x),\ldots,f_J(x)$ for small $J<K$, playing as main principal functions. 
Hence, it would be better to set smaller values of $\tau_j^2$ for small value of $j$.  

\subsection{Adaptation of orthogonality constraint}
While the proposed prior can be applied with fixed values of $\tau_k^2$, specific choice of these hyperparameters can significantly influence the resulting posterior inference. In practice, strict orthogonality (i.e., $\tau_k^2 = 0$) may be desirable for leading functions to ensure interpretability, whereas milder constraints (i.e., larger $\tau_k^2$) for higher-order functions allow more flexibility and help capture finer structures in the data.

To accommodate data-driven adaptation of these constraints, we introduce a hierarchical approach by assigning prior distributions to $\tau_k^2$. 
Specifically, we consider two strategies: a global prior and a local prior. 
In the global prior approach, we assume a common strength of constraint across all components, that is, $\tau_k^2 = \tau^2$ for all $k$, and assign an inverse-gamma prior $\tau^2 \sim \mathrm{IG}(a_0, b_0)$ with fixed hyperparameters $(a_0, b_0)$. This specification enforces a uniform level of orthogonality and is suitable when the same degree of orthogonality is desired across all functions.
In contrast, the local prior approach allows component-specific adaptation by assuming $\tau_k^2 \sim \mathrm{IG}(a_0, b_0)$ independently for each $k$. This formulation enables greater flexibility, allowing stronger constraints on leading functions while relaxing constraints on higher-order functions, which can be advantageous when capturing complex variations in the data. 
Because of such adaptivity of strength of orthogonal constraints, we call the prior {\it Adaptively Orthogonal (AO) prior}. 

From a computational perspective, introducing priors for $\tau_k^2$ only adds an additional inverse-gamma update step in the Gibbs sampler, preserving the overall simplicity of the posterior computation. 
Regarding the hyperparameters $(a_0, b_0)$, we use the default setting $a_0 = 3$ and $b_0 = 2/K^2$, which implies that the prior mean of $\tau^2$ is $\E[\tau^2] = 1/K^2$. This choice provides a weakly informative prior that favors moderate orthogonality (as $\tau^2$ is shrunk around a small value), while still allowing sufficient flexibility to adapt to the data.

\subsection{Properties of the AO prior}

In this section, we investigate theoretical properties of the proposed AO prior and clarify the effect of its hyperparameters. 
For simplicity, we assume that $\Omega=I_L$, that is, the basis functions are orthonormal.
Furthermore, we consider the choice of hyperparameter as $b_{0,j}=0_{L-j}$ and $B_{0,j}=\gamma I_{L-j}$ for some $\gamma>0$, and assume that the $(L-j)\times L$ matrix $H_{j+1}$ is orthogonal, that is, $H_{j+1}H_{j+1}^\top=I_{L-j}$. 
Under this setting, we consider a conditional expectation $\E[\|\beta_{j+1}\|^2| \beta_{1:j}]$, 
a conditional expectation of the squared norm of the coefficient vector of $f_{j+1}(x)$.
It follows that 
$$
A_{j+1}A_{j+1}^{\top}=
\left(\begin{array}{cc}
B_{1:j}^\top B_{1:j} &  (H_{L-j}B_{1:j})^{\top}\\
H_{L-j}B_{1:j} &  I_{L-j}
\end{array}\right)
$$
and 
$$
(A_{j+1}A_{j+1}^{\top})^{-1}=
\left(\begin{array}{cc}
(B_{1:j}^\top P_j B_{1:j})^{-1} &  -(B_{1:j}^\top P_j B_{1:j})^{-1}B_{1:j}^{\top}H_{L-j}^\top\\
-H_{L-j}B_{1:j}(B_{1:j}^\top P_j B_{1:j})^{-1} &  I_{L-j}+H_{L-j}B_{1:j}(B_{1:j}^\top P_j B_{1:j})^{-1}B_{1:j}^\top H_{L-j}^\top
\end{array}\right),
$$
where $B_{1:j}=(\beta_1,\ldots,\beta_j)$ and $P_j=I_L-H_{L-j}^\top H_{L-j}$.
Then, we have
\begin{align*}
{\rm tr}&[{\rm Var}(\beta_{j+1}|\beta_{1:j})]\\
&=
{\rm tr}\left[
{\rm blockdiag}(\tau_{j+1}^2 I_{j}, \gamma I_{L-j})(A_{j+1}A_{j+1}^{\top})^{-1}
\right]\\
&=\tau_{j+1}^2 {\rm tr}[(B_{1:j}^\top P_j B_{1:j})^{-1}\} + \gamma(L-j) + \gamma \ {\rm tr}\left\{(B_{1:j}^\top P_j B_{1:j})^{-1}B_{1:j}^\top  H_{L-j}^\top (I_L-P_j) H_{L-j} \right].
\end{align*}
Since $\E[\beta_{j+1}|\beta_{1:J}]=0_L$ from (\ref{eq:cond_prior}), the above expression is equivalent to $\E[\|\beta_{j+1}\|^2| \beta_{1:j}]$.
The conditional expectation of the squared norm (conditional variance) is an increasing function of $\tau_j^2$, indicating that weaker orthogonal conditions lead to larger conditional variance. 
Also, under fixed $\tau_j^2$ and $\gamma$, the conditional variance decreases as $j$ increases, showing that the coefficient $\beta_j$ for the $j$th function will be more concentrated around the origin for larger $j$. 
Such property would be preferable since the $j$th function with large $j$ would not have much contribution.

\begin{figure}[htbp!]
\centering
\includegraphics[width=\linewidth]{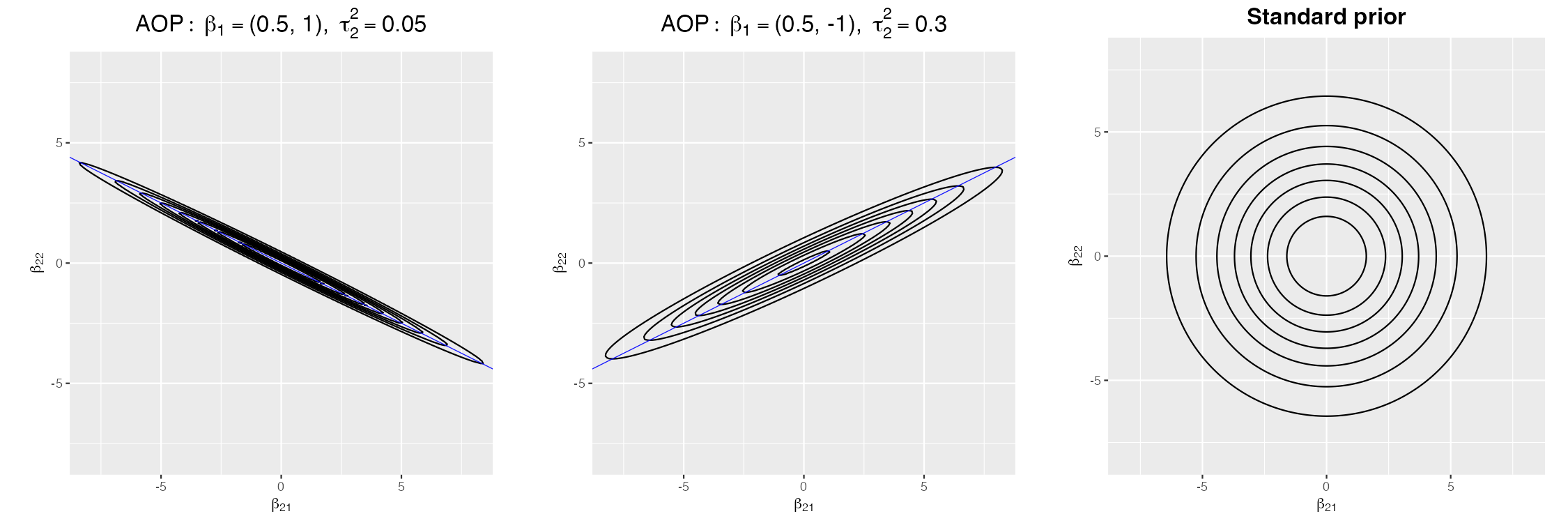}
\caption{Contour plots of AO prior for $\beta_2$ given $\beta_1$ with $\beta_1=(0.5, 1), \tau_2^2=0.05$ (left) and $\beta_1=(0.5, -1), \tau_2^2=0.3$ (center) and the standard prior, $\beta_2\sim N(0, 10I_2)$ (right). 
The solid blue line in right and center panels corresponds to the parameter space of $\beta_2$ where the orthogonal condition holds exactly. } 
\label{fig:prior}
\end{figure}

To visualize the proposed prior, we set $L\!=\!K\!=\!2$ and consider the conditional prior for $\beta_2$ given $\beta_1$. 
In Figure~\ref{fig:prior}, we show the contours of the conditional densities under $\beta_1=(0.5, 1), \tau_2^2=0.05$ and $\beta_1=(0.5, -1), \tau_2^2=0.3$, and that of the standard normal prior, $\beta_2\sim \Nc(0, 10I_2)$ for comparison, where $b_{02}=(0,0)^\top$ and $B_{02}=2I_2$ for the proposed prior. 
We observe that the proposed conditional prior has significant prior mass around the parameter space where the orthogonal constraint is exactly satisfied, i.e. $\{\beta_2=(t, -t/2), \ \ t\in (-\infty,\infty)\}$ for $\beta_1=(0.5, 1)$ and $\{\beta_2=(t, -t/2), \ t\in (-\infty,\infty)\}$  $\beta_2=\{(t, t/2), \ t\in (-\infty,\infty)\}$.
Moreover, the degree of concentration is dictated by the scale parameter $\tau_2$, so the smaller value of $\tau_2$ leads to more concentration on the orthogonal parameter space.  
Such properties are quite different from the standard independent prior, where the information of $\beta_1$ is not reflected to the prior of $\beta_2$. 
These visualizations highlight how the proposed prior adaptively balances orthogonality and variability through its hyperparameters.

\subsection{Posterior computation}

We provide a posterior computation algorithm for the proposed prior.
From the conditional prior specification (\ref{eq:cond_prior}), we first note that the joint prior of $\beta_1,\ldots,\beta_K$ can be obtained as 
\begin{align*}
&\prod_{j=1}^{K-1}|{\rm det}(A_{j+1}(\beta_{1:j}))|(\tau_{j+1}^2)^{-j/2}\exp\left[-\frac1{2\tau_{j+1}^2}\sum_{k=1}^{j}(\beta_k^\top \Omega \beta_{j+1})^2\right]\\
&\times \prod_{j=1}^K\exp \left[- \frac12(H_{j}\beta_{j}-b_{0j})^{\top} B_{0j}^{-1}(H_{j}\beta_{j}-b_{0j})\right],
\end{align*}
where $H_1=I_L$. 
Then, the full conditional prior of $\beta_k$ is proportional to 
\begin{align*}
&\prod_{j=k}^{K-1} |{\rm det}(A_{j+1}(\beta_{1:j}))|\exp\left[-\frac1{2\tau_{k}^2}\sum_{j=1}^{k-1}(\beta_j^\top \Omega \beta_k)^2 -\frac12\sum_{j=k+1}^K \frac{(\beta_j^\top \Omega \beta_k)^2}{\tau_j^2} \right] \\
&\times 
\exp\left[
-\frac12(H_{k}\beta_{k}-b_{0k})^{\top} B_{0k}^{-1}(H_{k}\beta_{k}-b_{0k})\right].
\end{align*}
Then, without the term $\prod_{j=k}^{K-1}|A_{j+1}(\beta_{1:j})|$, the above full conditional is a multivariate normal distribution $\Nc_L(A_{\beta_k}^{-1}B_{\beta_k}, A_{\beta_k}^{-1})$, where 
\begin{align*}
&A_{\beta_k}=H_k^\top B_{0k}^{-1} H_k + \frac1{\tau_{k}^2}\sum_{j=1}^{k-1}\Omega \beta_j\beta_j^\top \Omega + \sum_{j=k+1}^K \frac{ \Omega \beta_j\beta_j^\top \Omega}{\tau_j^2}, \ \ \ \ \ 
B_{\beta_k}=H_k^\top B_{0k}^{-1}b_{0k}\,.
\end{align*}
Hence, when the likelihood function for $\beta_k$ is Gaussian, we can use a simple Metropolis-Hastings algorithm via generating proposals from the normal distribution. 
The specific example will be given in Section~\ref{sec:FPCA}.

\subsection{Sparsity-inducing prior}\label{sec:sparse}
Since the AO prior is a valid multivariate prior distribution, we can introduce sparsity-inducing prior into the AO prior.
Specifically, we adopt a hierarchical prior of the form, $H_k\beta_k\sim \mathcal{N}_{L-k+1}(0, \gamma_k I_{L-k+1})$ and $\sqrt{\gamma_k}\sim {\rm HC}_+$, where ${\rm HC}_+$ denotes the standard half-Cauchy prior.
This corresponds to setting $B_{0k}=\gamma_k I_{L-k+1}$ in the general construction of the AO prior and this formulation leads to the horseshoe prior \citep{carvalho2009handling} for the marginal prior of $H_k\beta_k$. 
When $H_k\beta_k$ is close to the origin, the $k$th function is $f_k(x)\approx 0$, so that irrelevant functions could be automatically eliminated in the posterior.
To perform the posterior computation with the sparsity-inducing prior, we can use the sampling algorithm in the previous section, under given $\gamma_k$.
To generate random sample of $\gamma_k$, we introduce an additional variable $\eta_k$ to represent $\gamma_k|\eta_k\sim {\rm IG}(1/2, 1/\eta_k)$ and $\eta_k\sim {\rm IG}(1/2, 1)$, which gives $\sqrt{\gamma_k}\sim {\rm HC}_+$ as the marginal distribution. 
Then, sampling from the full conditional of $\gamma_k$ is performed by two steps.
We first generate $\eta_k$ from ${\rm IG}(1, 1+1/\gamma_k)$ and then generate $\gamma_k$ from ${\rm IG}(\widetilde{a}_k, \widetilde{b}_k)$, where 
$$
\widetilde{a}_k=\frac12 + \frac{L-k}{2},\ \ \ \ \ \widetilde{b}_k=\frac12(H_{k}\beta_{k}-b_{0k})^{\top} (H_{k}\beta_{k}-b_{0k}) + \frac{1}{\eta_k}.
$$

\section{Bayesian functional principal component analysis}\label{sec:FPCA}

We demonstrate the usefulness of the proposed prior in the Bayesian functional principal component analysis (FPCA).
Given the functional observations, $X_1(t),\ldots,X_n(t)$ on $t\in \mathcal{T}$, the Bayesian FPCA assumes the following low-rank model: 
\begin{equation}\label{eq:FPCA}
X_i(t)=\sum_{k=1}^K Z_{ik}f_k(t) + \ep_i(t), \ \ \ \ i=1,\ldots,n,
\end{equation}
where $f_k(t)$ is a principal function common to all the observations, $Z_{ik}$ is an observation-specific random weight for that function and $\ep_i(t)$ corresponds to an error term. 
We assume that $Z_{ik}\sim \Nc(0, \lambda_k)$, where $\lambda_k$ is an unknown parameter satisfying $\lambda_1>\cdots>\lambda_K$. 
Let $t_{i1},\ldots,t_{im_i}$ be a set of observed points for the $i$th observation. 
By introducing basis function representation for the principal function $f_k(x)$, Bayesian inference on FPCA is considered in \cite{suarez2017bayesian} and \cite{nolan2025bayesian}, and its robust version is also discussed in \cite{zhang2025robust}.
However, their prior distributions for $f_k(x)$ does not guarantee orthogonality each other, which may lead to inefficient low-rank representations.
We here introduce the proposed orthogonal prior for a sequence of functions, $f_1(x),\ldots,f_K(x)$, to obtain (nearly) orthogonal posterior inference on the principal functions. 

We perform posterior inference using a Gibbs sampler, which iteratively updates each parameter from its full conditional distribution. 
The detailed steps are described as follows:

\begin{itemize}
\item[-]\ (Sampling of $\beta_k$) \ \ For $k=1,\ldots,K$, generate a proposal $\beta_k^{\ast}$ of $\beta_k$ from $\Nc_L(V_k^{-1}U_k, V_k^{-1})$, where
\begin{equation*}
\left\{\!\!
\begin{array}{ll}
 &\displaystyle V_k = H_k^\top B_{0k}^{-1} H_k 
+ \frac{1}{\tau_k^2}\sum_{j=1}^{k-1}\Omega \beta_j \beta_j^\top \Omega 
+ \sum_{j=k+1}^{K}\frac{\Omega \beta_j \beta_j^\top \Omega}{\tau_j^2} 
+ \frac{1}{\sigma^2}\sum_{i=1}^{n} Z_{ik}^2 \Phi_i \Phi_i^\top,\\
 &\displaystyle U_k = H_k^\top B_{0k}^{-1} b_{0k} + \frac{1}{\sigma^2}\sum_{i=1}^{n} Z_{ik}\left(X_i - \sum_{l \neq k}Z_{il} F_{il}\right).
\end{array}
\right.
\end{equation*}
Here, $\Phi_i=(\phi(t_{i1}),\ldots,\phi(t_{im_i}))^\top $ is the $m_i$-dimensional vector of basis functions evaluated at the observed points, and $F_{il}=(f_l(t_{i1}),\ldots,f_l(t_{im_i}))^\top$ denotes the principal function evaluated at the observed points. 
Then, accept the proposal $\beta_k^{\ast}$ with probability 
$$
\min\bigg(1, \prod_{j=k}^{K-1}\frac{|{\rm det}(A_{j+1}(\beta_{1:j}^{\ast}))|}{|{\rm det}(A_{j+1}(\beta_{1:j}))|}\bigg),
$$
where $\beta_{1:j}^{\ast}$ is defined by replacing $\beta_k$ in $\beta_{1:j}$ with $\beta_k^{\ast}$.

\item[-] \ (Sampling of $Z_{ik}$) \ \ 
For $i = 1,\ldots,n$ and $k = 1,\ldots,K$, draw $Z_{ik}$ from its full conditional distribution, $\Nc(\mu_{ik}^{\ast}, v_{ik}^{\ast})$, 
where
$$
v_{ik}^{\ast}
= \left(\frac{F_{ik}^\top F_{ik}}{\sigma^2} + \frac{1}{\lambda_k}\right)^{-1}, \quad
\mu_{ik}^{\ast} = \frac{v_{ik}^{\ast} F_{ik}^\top}{\sigma^2} \left(X_i - \sum_{l \neq k} Z_{il}F_{il}\right).
$$

\item[-]\ (Sampling of $\lambda_k$) \ \ For $k=1,\ldots,K$, draw $\lambda_k$ from its full conditional distribution, ${\rm IG}\left(a_{\lambda} + n/2, \, b_{\lambda} + \sum_{i=1}^{n} Z_{ik}^2/2 \right)$.

\item[-] \ (Sampling of $\tau_k^2$) \ \ For $k=1,\ldots,K$, draw $\tau_k^2$ from its full conditional distribution, $\sim {\rm IG}\left(a_0 + (k-1)/2, \, b_0 + \sum_{j=1}^{k-1} (\beta_j^\top \Omega \beta_k)^2/2 \right)$.

\item[-] \ (Sampling of $\sigma^2$) \ \ Draw $\sigma^2$ from its full conditional distribution, ${\rm IG}(a_\sigma+\sum_{i=1}^nm_i/2, b_\sigma + \sum_{i=1}^n\sum_{j=1}^{m_i}(X_{ij}-\sum_{k=1}^KZ_{ik}f_{k}(t_{ij}))^2/2)$.
\end{itemize}

As shown above, all these full conditional distributions are standard distributions, which allows for efficient posterior computation.

\section{Simulation}\label{sec:sim}

\subsection{Illustration of the AO prior}
We conducted simulation experiments to evaluate the performance of the proposed orthogonality-inducing priors in the context of functional principal component analysis (FPCA) described in Section~\ref{sec:FPCA}.
We considered two scenarios for the underlying true principal functions: (i) Legendre polynomials and (ii) Haar wavelets. For both scenarios, functional observations were generated at $T = 30$ equally spaced time points on $[0,1]$, and the number of samples $n$ was set to 50, 100, or 200.  
In scenario (i), the first three true principal functions were defined as scaled Legendre polynomials, and in scenario (ii), as scaled Haar wavelet functions, described as
\begin{align*}
f_1(t) &= \sqrt{3}\,\bigl(2t - 1\bigr), ~
f_2(t) = \sqrt{5}\,\bigl(6t^2 - 6t + 1\bigr), ~
f_3(t) = \sqrt{7}\,\bigl(20t^3 - 30t^2 + 12t - 1\bigr)
\end{align*}
for the Legendre polynomial scenario, and
\begin{align*}
f_1(t) &= \I\Big(0 \le t < \frac12\Big) - \I\Big(\frac12 \le t \le 1\Big), \ \ \ \ 
f_2(t) = \sqrt{2} \left[ \I\Big(0 \le t < \frac14\Big) - \I\Big(\frac14 \le t < \frac12\Big) \right], \\
& \hspace{2.7cm}
f_3(t) = \sqrt{2} \left[ \I\Big(\frac12 \le t < \frac34\Big) - \I\Big(\frac34 \le t \le 1\Big) \right].
\end{align*}
for the Haar wavelet scenario, where $\I(\cdot)$ is an indicator function. 
The latent scores for the first three components were generated independently from normal distributions with decreasing variances to reflect dominant leading components: $Z_{i1} \sim \Nc(0,1)$, $Z_{i2} \sim \Nc(0,0.7^2)$, and $Z_{i3} \sim \Nc(0,0.5^2)$. The observed functional data $Y_i(t)$ were then obtained by adding Gaussian noise with standard deviation $\sigma = 1$ to the true signal.

For the generated dataset, we applied the proposed adaptive orthogonal prior with a global inverse-gamma prior on $\tau^2$ (we call AO-G), and local inverse-gamma priors on each $\tau_k^2$ (AO-L), where cubic B-splines with an intercept are used for basis functions. 
For both priors, we incorporate the sparsity-inducing prior described in Section ~\ref{sec:sparse}.
For comparison, we adopted a non-orthogonal prior (NO) which uses a standard normal prior for each coefficients, and a non-orthogonal shrinkage prior (NO-S) applying the horseshoe prior \citep{carvalho2009handling} for coefficients of basis functions to reduce the number of irrelevant principal functions.
Moreover, we compared nearly mutually orthogonal (NeMO) processes \citep{matuk2022bayesian}. 
The (largest) number of possible components was set to $K=10$ in all the methods, and the number of basis functions in NO, NO-S, AO-G and AO-L was set to $L=12$, noting that the results were not sensitive to the choice of the different values for $L$.
For NeMO, the threshold value for orthogonal constraint is set to $10^{-4}$, following the default setting of the R package provided by the author's GitHub repository. 
Posterior inference was conducted via MCMC, running for 3,000 iterations after discarding 2,000 burn-in iterations.

\begin{figure}[t]
\centering
\includegraphics[width=\linewidth]{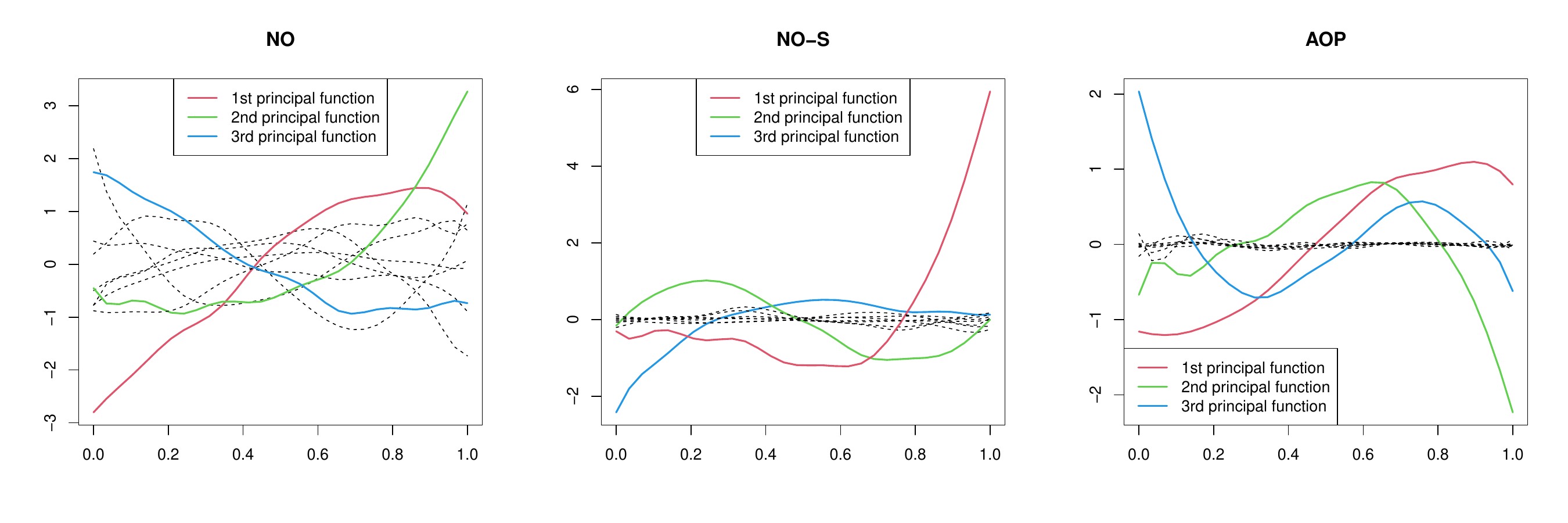}
\includegraphics[width=\linewidth]{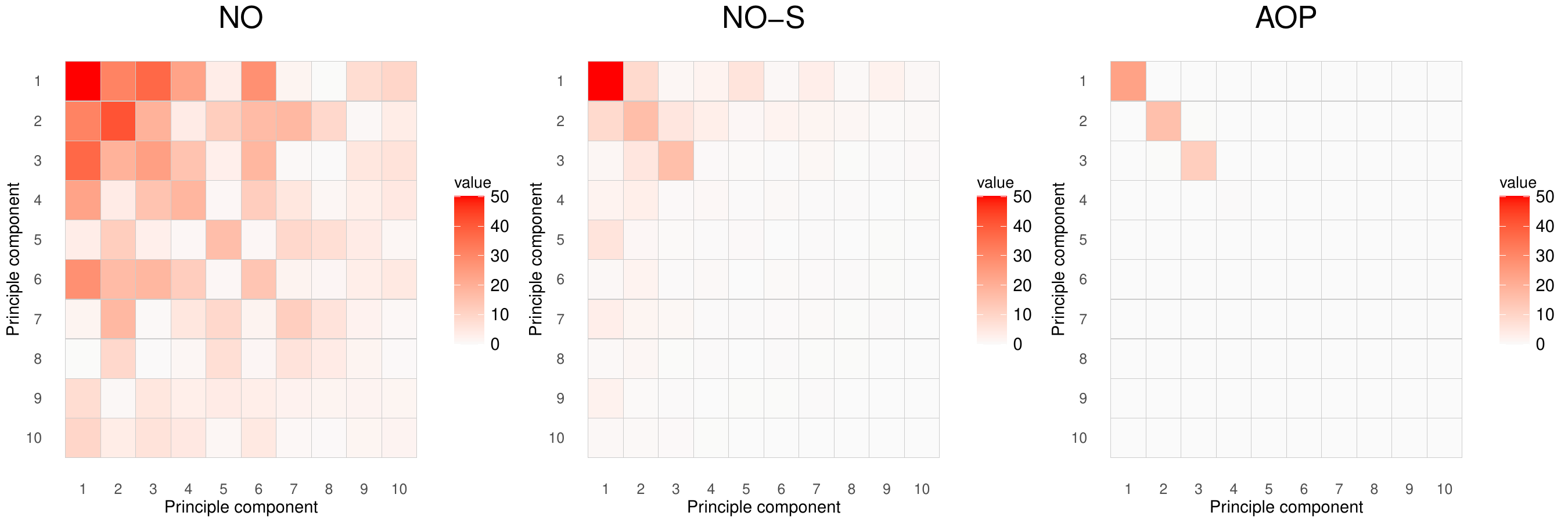}
\caption{ Estimated principal functions (upper) and heatmaps of inner products of principal functions (lower). } 
\label{fig:sim}
\end{figure}

We first show the result with a single simulation data under scenario (i). 
In Figure~\ref{fig:sim}, we visualize the estimated principal functions and heatmaps of their inner products, obtained by ``NO" (non-orthogonal) and ``AO-G" (orthogonal) methods.
Overall, the results demonstrated that our proposed orthogonal priors successfully enforced near-orthogonality, effectively selected the number of components, and improved reconstruction accuracy while maintaining reasonable uncertainty quantification. In particular, the orthogonal approaches accurately estimated the true number of principal components (three), with the remaining components being automatically shrunk toward zero. This shrinkage behavior is evident in the estimated principal functions, where only the leading three functions show substantial variation, while the others are effectively eliminated.
Furthermore, the heatmaps of inner products provide a clear contrast between the methods: under the proposed orthogonal priors, the off-diagonal elements are nearly zero, indicating that the estimated functions are nearly orthogonal. In contrast, the non-orthogonal methods exhibit large off-diagonal inner products, implying strong correlations among components and reducing interpretability. These findings highlight the advantage of orthogonality-inducing priors in achieving interpretable, parsimonious representations in functional principal component analysis.

\subsection{Comparison via Monte Carlo experiments}
We next evaluate the performance through Monte Carlo simulation studies. 
For each replication, we evaluated the following two metrics:
\begin{itemize}
\item[-] 
{\bf Number of effective components (NC)}: The number of components with sufficiently large variations, defined as ${\rm NC}=\sum_{k=1}^K \I\big\{\int \widehat{f}_k^2(t)dt>\epsilon\big\}$, where $\widehat{f}_k(t) \ (k=1,\ldots,K)$ is the posterior mean of the principal functions. 
We set $\epsilon=0.1$ in this study. 

\item[-]
{\bf Orthogonality measure (OG)}: The sum of absolute off-diagonal inner products among estimated principal functions, defined as ${\rm OG}=\sum_{k=2}^K \sum_{j=1}^{k-1} \Big|\int \widehat{f}_j(t)\widehat{f}_k(t)dt\Big|$.
Note that the smaller values indicate stronger orthogonality.

\item[-]
{\bf Mean squared error (MSE)}: Estimation accuracy of the mean function $\mu_i(t)=Z_{i1}f_1(t)+Z_{i2}f_2(t)+Z_{i3}f_3(t)$, defined as $n^{-1}\int \{\widehat{\mu}_i(t)-\mu_i(t)\}^2 dt$, where $\widehat{\mu}_i(t)$ is the posterior mean of the mean function.

\item[-]
{\bf Interval score (IS)}: Coverage performance of the point-wise $95\%$ credible intervals of the mean function $\mu_i(t)$, defined in \cite{gneiting2007strictly}.
\end{itemize}

Tables~\ref{tab:sim1} summarizes the averaged values of NC and OG, based on 200 Monte Carlo replications. 
It shows that the non-orthogonal model without shrinkage (NO) consistently overestimates the number of components, returning the maximum value of $K = 10$ in all cases. 
Although the non-orthogonal model with shrinkage (NO-S) reduces the number of estimated components, it still tends to slightly overestimate when the sample size is large. 
In contrast, the adaptive orthogonal prior approaches (AO-G and AO-L) provide NC values close to the true number of components, demonstrating the effectiveness of orthogonality-inducing priors in promoting sparsity.
On the other hand, NeMO tends to produce NC values larger than the true number of components. 
Regarding OG, the proposed AO-G and AO-L yield considerably smaller values than the other methods, indicating that the estimated principal functions are nearly orthogonal. 
On the other hand, both NO and NO-S exhibit large OG values, especially in large sample settings, reflecting strong dependencies among components. 
Also, it is observed that the OG values of NeMo are slightly larger than those of AO-G and AO-L.
These results confirm that the proposed priors not only improve parsimony but also enhance interpretability by ensuring near-orthogonal decomposition.

Table~\ref{tab:sim2} provides the averaged values of MSE and IS, except for NeMO.
The reason is that the MSE of NeMO is larger than the other method possibly due to the difference of underlying models (Gaussian process for NeMo and basis functions for the other methods), so that this would not provide fair comparison of the performance. 
The results show that the proposed AO-G and AO-L priors provides better estimation and inference performance than than non-orthogonal priors, indicating that incorporating adaptive orthogonality enhances not only interpretability but also accuracy of estimation and inference.  

\begin{table}[t]
\centering
\caption{Average values of estimated number of principal component functions (NC) and orthogonality measures (OG) among principal functions, based on 200 Monte Carlo replications. }
\label{tab:sim1}
\medskip
{\small 
\begin{tabular}{cccccccccccccccc}
\hline
&&& \multicolumn{5}{c}{Scenario 1} &&  \multicolumn{5}{c}{Scenario 2}\\
 & $n$  &  & {\footnotesize NO} & {\footnotesize NOS} & {\footnotesize NeMO} & {\footnotesize AO-G} & {\footnotesize AO-L} &  & {\footnotesize NO} & {\footnotesize NO-S} & {\footnotesize NeMO} & {\footnotesize AO-G} & {\footnotesize AO-L} \\
 \hline
 & 50 &  & 8.48 & 3.01 & 5.08 & 3.01 & 3.02 &  & 8.35 & 3.02 & 5.19 & 3.00 & 3.01 \\
{\footnotesize NC} & 100 &  & 8.94 & 3.03 & 5.43 & 3.00 & 3.00 &  & 8.65 & 3.06 & 5.34 & 3.00 & 3.00 \\
 & 200 &  & 9.16 & 3.28 & 5.34 & 3.00 & 3.00 &  & 8.94 & 3.30 & 5.32 & 3.00 & 3.00 \\
 \hline
 & 50 &  & 16.0 & 2.49 & 0.84 & 0.05 & 0.04 &  & 18.1 & 3.91 & 0.58 & 0.05 & 0.03 \\
{\footnotesize OG} & 100 &  & 23.2 & 3.02 & 0.87 & 0.05 & 0.02 &  & 24.3 & 3.80 & 0.58 & 0.05 & 0.02 \\
 & 200 &  & 30.2 & 4.46 & 0.85 & 0.05 & 0.03 &  & 30.5 & 5.38 & 0.56 & 0.04 & 0.03 \\
\hline
\end{tabular}
}
\end{table}

\begin{table}[t]
\centering
\caption{Average values of mean squared errors (MSE) of the posterior mean of the underlying functions and interval scores (IS) of the $95\%$ point-wise credible intervals of the underlying functions, based on 200 Monte Carlo replications. }
\label{tab:sim2}
\medskip
{\small 
\begin{tabular}{cccccccccccccccc}
\hline
&&& \multicolumn{4}{c}{Scenario 1} &&  \multicolumn{4}{c}{Scenario 2}\\
 & $n$  &  & {\footnotesize NO} & {\footnotesize NOS} & {\footnotesize AO-G} & {\footnotesize AO-L} &  & {\footnotesize NO} & {\footnotesize NO-S} &  {\footnotesize AO-G} & {\footnotesize AO-L} \\
 \hline
 & 50 &  & 0.16 & 0.13 & 0.13 & 0.13 &  & 0.16 & 0.14 &  0.13 & 0.13 \\
{\footnotesize MSE} & 100 &  & 0.13 & 0.12 & 0.11 & 0.11 &  & 0.13 & 0.12 &  0.12 & 0.12 \\
 & 200 &  & 0.11 & 0.11 &  0.11 & 0.11 &  & 0.12 & 0.11  & 0.11 & 0.11 \\
 \hline
 & 50 &  & 1.85 & 1.68 & 1.63 & 1.62 &  & 1.91 & 1.77 & 1.70 & 1.69 \\
{\footnotesize IS} & 100 &  & 1.68 & 1.56 &  1.54 & 1.52 &  & 1.76 & 1.66 &  1.62 & 1.61 \\
 & 200 &  & 1.57 & 1.49  & 1.47 & 1.46 &  & 1.65 & 1.58 &  1.56 & 1.55 \\
\hline
\end{tabular}
}
\end{table}

\subsection{Sensitivity of the scale parameter for constraint}

To check the sensitivity of the scale parameter $\tau^2$ controlling the strength of the orthogonality in the proposed AO prior, we evaluated the performance with several choices of $\tau^2$. 
Using the same data generating scenarios, we evaluated the performance by the four evaluation values introduced in the previous section.
The results are presented in Table~\ref{tab:sim-tau-effect}.
It highlights that simply fixing the scale parameter $\tau^2$ at a very small value does not necessarily lead to improved performance. 
While a small $\tau^2$ enforces strong orthogonality in the prior, overly stringent constraints can introduce numerical instability and reduce flexibility, which may deteriorate data fit and result in larger MSE, unnecessarily large number of non-empty functions or inefficient posterior inference (i.e. larger IS). 
On the other hand, setting $\tau^2$ too large weakens orthogonality constraints as expected. 
As a consequence, the estimated functions may exhibit non-orthogonality as confirmed by larger values of OG. 
In contrast, the proposed AO-G approach, which estimates $\tau^2$ adaptively from the data, is able to automatically learn an appropriate balance between orthogonality and flexibility.
The resulting performance demonstrates that adaptive learning of the constraint strength is crucial for achieving both accurate estimation and effective orthogonality, underscoring the necessity of treating $\tau^2$ as an inferential quantity rather than a fixed tuning parameter.

\begin{table}[t]
\centering
\caption{Performance of the orthogonal prior with fixed $\tau$ and AO-G prior, based on 200 Monte Carlo replications. }
\label{tab:sim-tau-effect}

\medskip
\begin{tabular}{ccccccccccccccc}
\hline
&&& \multicolumn{6}{c}{$\tau^2$} & \\
 & $n$ &  & $10^{-4}$ & $10^{-3}$ & $0.01$ & $0.05$ & $0.2$ & $0.5$ & AO-G \\
\hline
 & 50 &  & 5.35 & 5.17 & 4.86 & 3.37 & 3.00 & 3.00 & 3.00 \\
NC & 100 &  & 6.21 & 6.26 & 5.36 & 3.26 & 3.00 & 3.00 & 3.00 \\
 & 200 &  & 7.00 & 7.05 & 5.45 & 3.13 & 3.00 & 3.00 & 3.00 \\
 \hline
 & 50 &  & 0.10 & 0.11 & 0.12 & 0.08 & 0.05 & 0.09 & 0.06 \\
OG & 100 &  & 0.11 & 0.10 & 0.13 & 0.06 & 0.04 & 0.08 & 0.05 \\
 & 200 &  & 0.10 & 0.10 & 0.13 & 0.04 & 0.04 & 0.07 & 0.05 \\
 \hline
 & 50 &  & 0.20 & 0.21 & 0.18 & 0.14 & 0.13 & 0.13 & 0.13 \\
MSE & 100 &  & 0.22 & 0.22 & 0.19 & 0.12 & 0.11 & 0.12 & 0.11 \\
 & 200 &  & 0.23 & 0.23 & 0.18 & 0.11 & 0.11 & 0.11 & 0.11 \\
 \hline
 & 50 &  & 2.02 & 2.12 & 1.91 & 1.67 & 1.63 & 1.66 & 1.63 \\
IS & 100 &  & 2.09 & 2.10 & 1.92 & 1.57 & 1.54 & 1.57 & 1.54 \\
 & 200 &  & 2.16 & 2.16 & 1.90 & 1.49 & 1.48 & 1.50 & 1.48 \\
\hline
\end{tabular}
\end{table}

\section{Data Analysis}

\begin{figure}[t]
\centering
\includegraphics[width=0.9\linewidth]{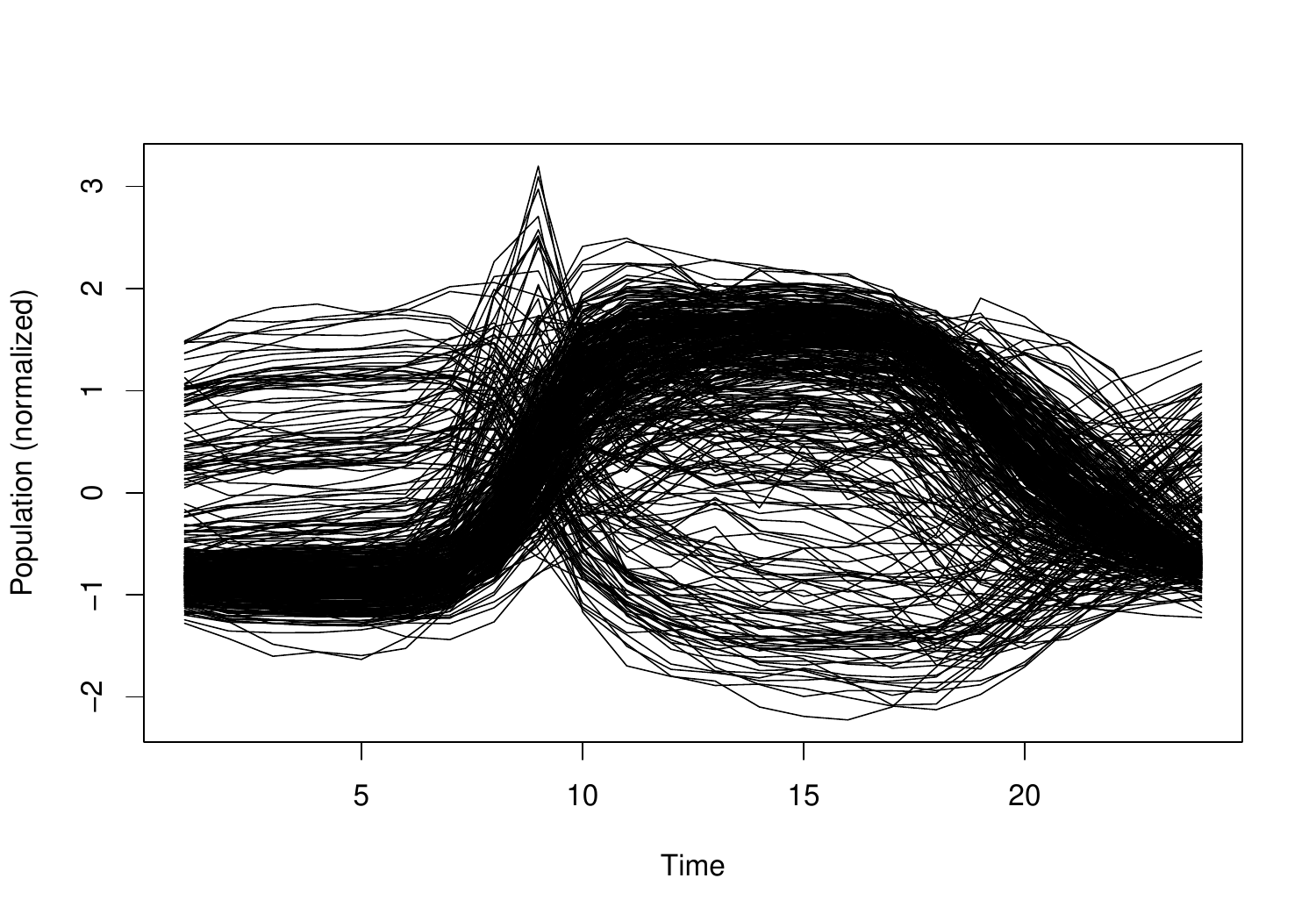}
\caption{ Scaled functional observations ($n=452$) of populations in Tokyo area.} 
\label{fig:app-data}
\end{figure}

Here we demonstrate the usefulness of the proposed orthogonal prior in Bayesian FPCA, using the population data collected by NTT Docomo Inc., which is the predominant mobile company in Japan.
Following the previous studies \citep{wakayama2025similarity,wakayama2024spatiotemporal}, we focus on the seven special wards of Tokyo metropolitan area and 500-meter mesh grids covering these areas.
While the original data were provided as three-dimensional arrays, representing locations, time points, and dates, we use the aggregated data consisting of average hourly population counts in February 2019 on weekdays, which results in $n=452$ functional observations with $m_i\equiv m=24$ discrete observed hours.
For $i=1,\ldots,n$, let $Y_i^{\ast}(t)$ be the aggregated functional observation, and we use the scaled version $Y_i(t)$ defined as $Y_i(t)=Y_i^{\ast}(t)/\sqrt{m^{-1}\sum_{j=1}^{m} Y_i^{\ast}(t_j)^2}$. 
Figure~\ref{fig:app-data} shows the scaled functional data, which clearly exhibits patterns indicate strong temporal variation, which motivates the use of a low-rank functional representation via FPCA.

\begin{figure}[b!]
\centering
\includegraphics[width=0.9\linewidth]{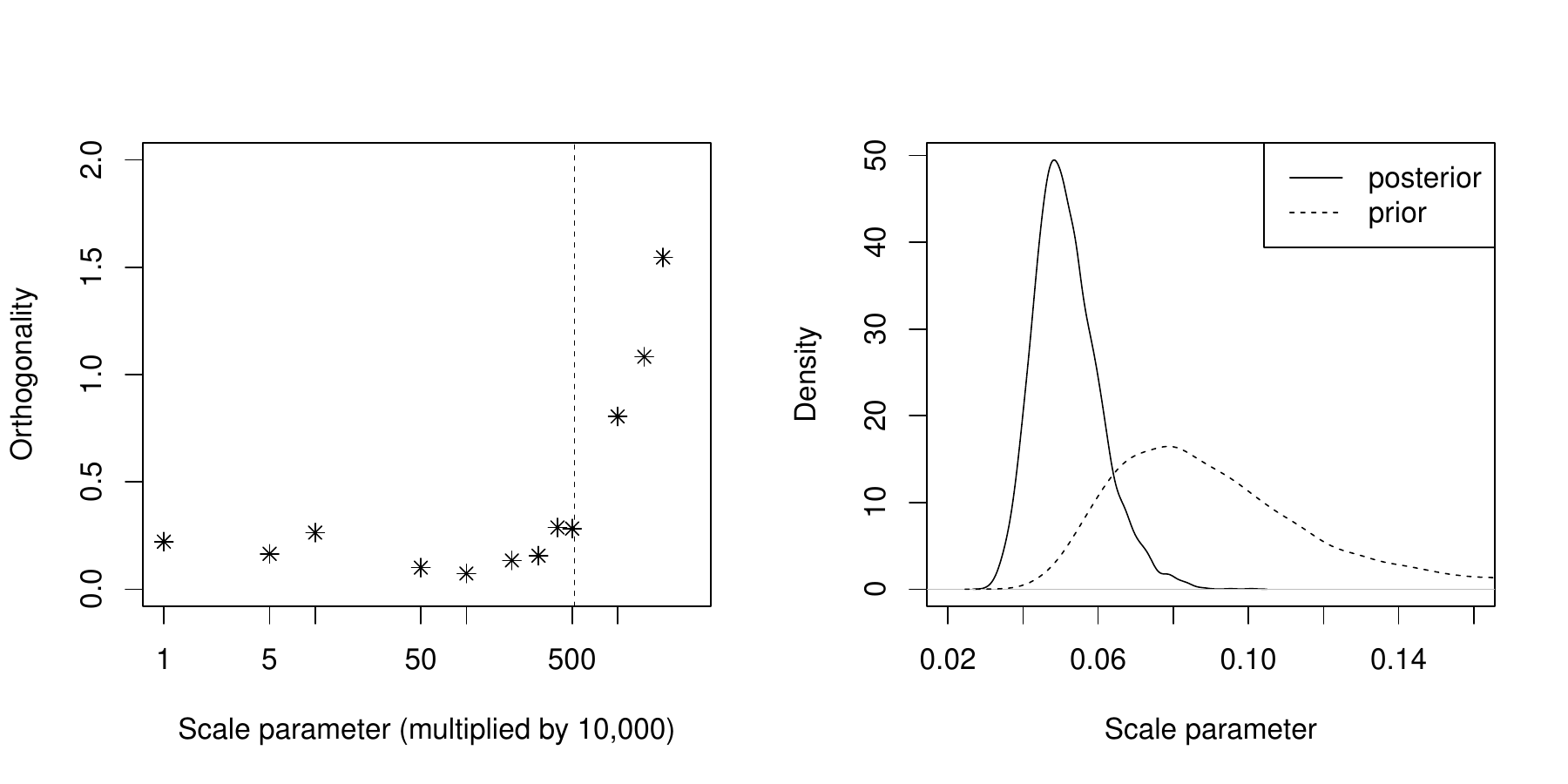}
\caption{ Orthogonality measures as a function of scale parameter $\tau$ (left) and posterior and prior densities (right) of the scale parameter $\tau$ to control the strength of orthogonality in AO-G. 
The dotted line in the left panel indicates the posterior mean of $\tau$. 
} 
\label{fig:app-tau}
\end{figure}

We fitted the Bayesian FPCA with four priors, NO, NO-S, NeMO and AO-G, to these functional data.
We adopted the same settings for the prior distributions and tuning parameters as used in Section~\ref{sec:sim}, such as $K = 10$ (maximum number of principal functions) and $L=12$ (the number of basis functions).
We generated 5,000 iterations after 5,000 burn-in periods.
In Figure~\ref{fig:app-tau}, we present the posterior distribution of $\tau$ (scale parameter controlling the degree of orthogonality constraint) in AO-G, and the orthogonality measures for fixed $\tau$. 
These results show that $\tau$ can be successfully learned from data and its posterior mean is not extremely small, which is a different feature from that of NeMo imposing nearly orthogonal constraint. 
We observe that the orthogonality measure remains below a stable threshold for values of $\tau$ up to approximately $0.05$, while it increases monotonically for larger values of $\tau$. 
Notably, the posterior mean of $\tau$ under the proposed method is located around this boundary.
Based on the simulation results in Section~\ref{sec:sim}, such choice of $\tau$ would be reasonable to balance orthogonality and efficiency.  

\begin{figure}[p]
\centering
\includegraphics[width=\linewidth]{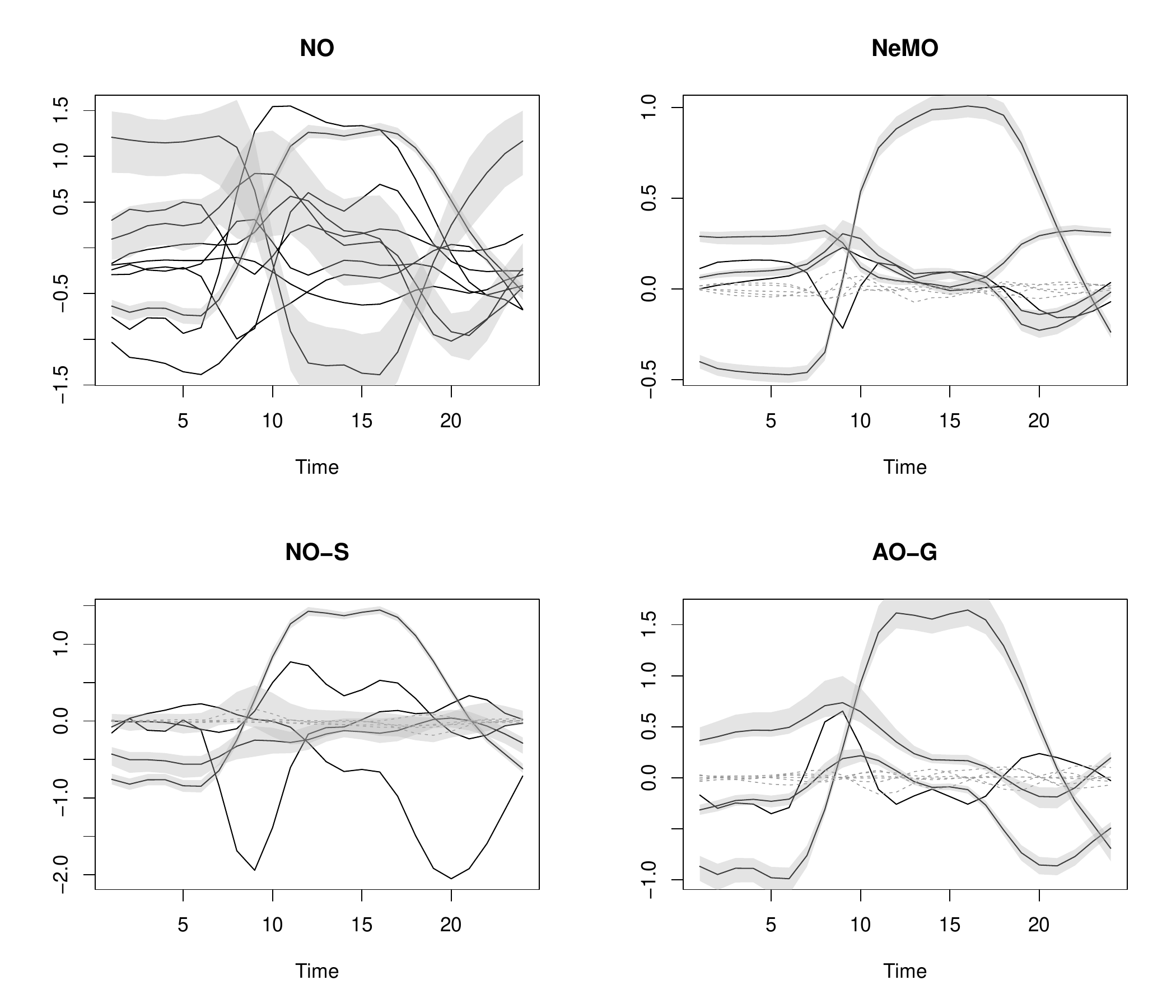}
\caption{ Estimated principal functions, where principal functions with small variations are expressed by dotted grey lines.  
}
\label{fig:app-func}
\end{figure}

\begin{figure}[htbp!]
\centering
\includegraphics[width=\linewidth]{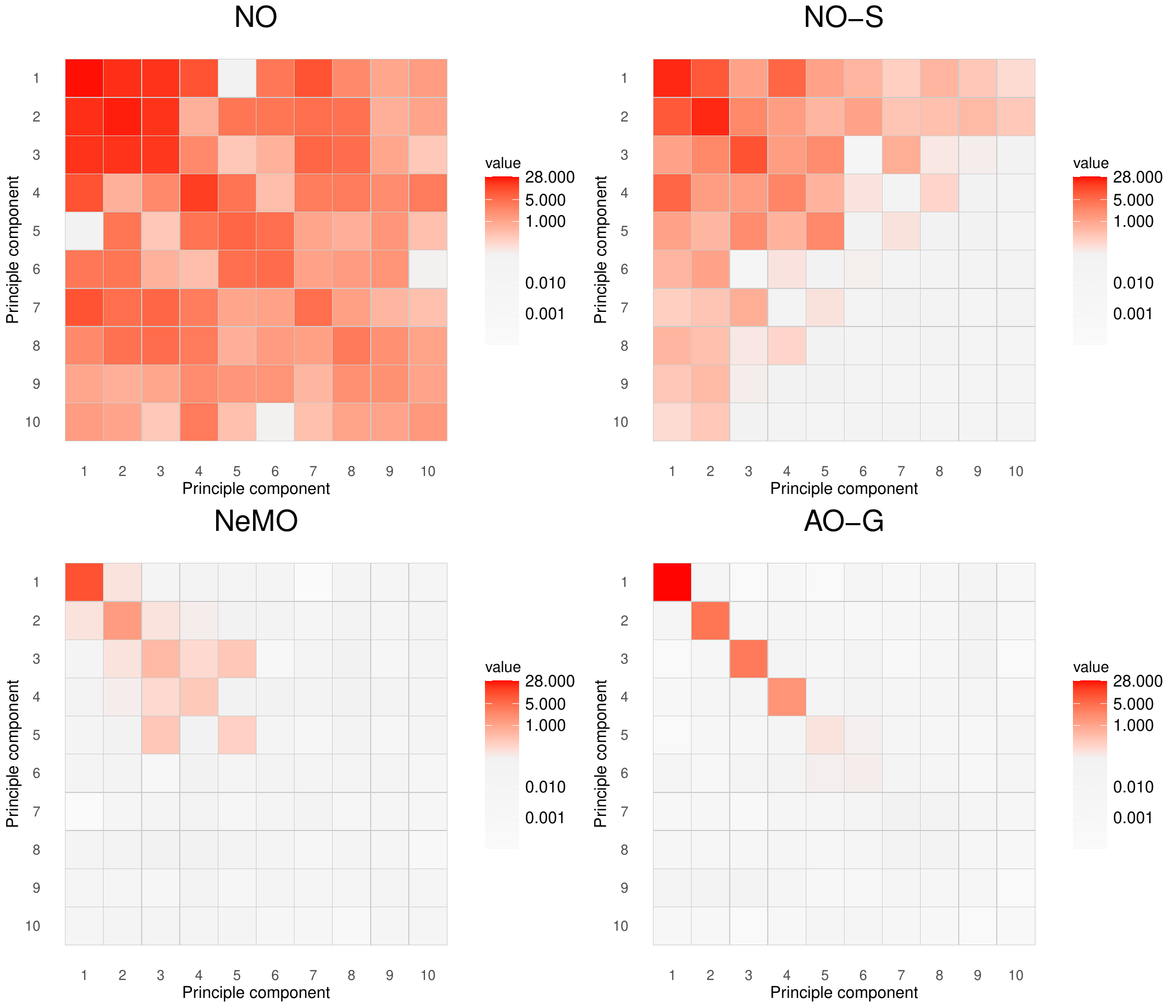}
\caption{ The heatmaps of inner products of principal functions.  
}
\label{fig:app-ortho}
\end{figure}

Figure~\ref{fig:app-func} displays the posterior means of the principal functions for each method, where only principal functions having variations $\int \widehat{f}_k(t)^2dt > 0.2$ are displayed by solid lines.
The number of such principal functions are 10 in NO, 6 in NO-S, 5 in NeMO and 4 in AO-G. 
Also, we present inner products of the estimated principal functions in Figure~\ref{fig:app-ortho}.
From these figures, it is observed that the proposed AO-G method yields interpretable leading components, effectively capturing dominant temporal variations such as morning and evening population peaks.
Notably, the orthogonal priors automatically eliminate unnecessary components by shrinking their amplitudes toward zero, resulting in a sparse low-rank representation that clearly highlights the dominant modes of variation. 
In contrast, the non-orthogonal approaches (NO and NO-S) suffer from overlapping principal components even when shrinkage prior is applied. 
While the NeMO method provides more orthogonal principal functions than NO and NO-S, the first principal function is dominant and the other significant principal functions are relatively similar, compared with those of AO-G. 
In fact, the orthogonality measures used in Section~\ref{sec:sim} are 1.72 (NeMO) and 0.18 (AO-G), which would indicate that AO-G yields more orthogonal principal functions. 

\begin{figure}[htbp!]
\centering
\includegraphics[width=0.9\linewidth]{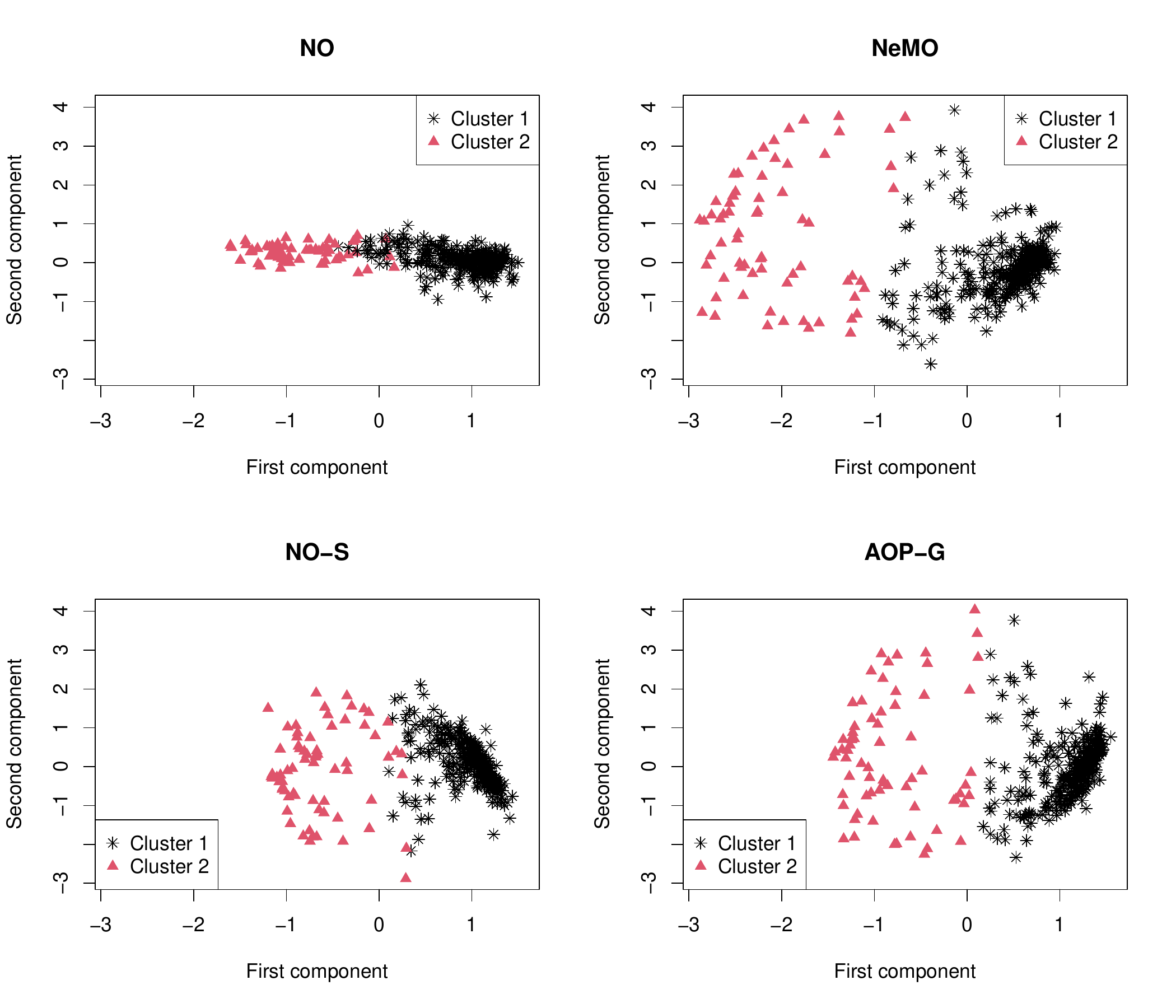}
\caption{ Scatter plots of the loadings of the first and second principal functions, corresponding to $(Z_{i1},Z_{i2})$ in the FPCA (\ref{eq:FPCA}), obtained from the four methods. } 
\label{fig:app-loadings}
\end{figure}

To further investigate the usefulness of the proposed methods, we projected the functional observations into two dimensional principal component score, $(Z_{i1},Z_{i2})$, which are shown in Figure~\ref{fig:app-loadings}. 
The colors in the scatter plots represent two groups obtained via two-class $k$-means clustering on the original space (black and red). 
We observe that the orthogonal approaches, NeMO and AO-G result in clearer separation of clusters than the non-orthogonal approaches, NO and NO-S. 
Further, it can be seen that two clusters are successfully separated only by the first principal component function detected by AO-G. 
This would exhibit the advantage of AO-G to extract more meaningful principal functions and provide efficient low-rank representations for functional data.


\section{Concluding Remarks}

In this paper, we have proposed a novel class of prior distributions for sequences of orthogonal functions, motivated by applications in functional data analysis and beyond. Our approach introduces a hierarchical construction that sequentially defines function coefficients via conditionally Gaussian distributions with approximate orthogonality constraints. By controlling these constraints through hyperparameters, our framework offers a flexible balance between strict orthogonality and modeling flexibility.
Despite these promising results, there remain several directions for future research. 
While we focused on basis function representations with a finite number of basis, the framework could be potentially extended to Gaussian process models, given the equivalence between GP priors and basis function representations with random coefficients. 
Such extensions would provide an adaptive version of the constraint prior developed by \cite{matuk2022bayesian}.

\section*{Acknowledgement}
This work is partially supported by JSPS KAKENHI Grant Numbers 24K21420 and 25H00546.

\vspace{1cm}
\bibliographystyle{chicago}
\bibliography{ref}

\begin{thebibliography}{}

\bibitem[\protect\citeauthoryear{Bai}{Bai}{2009}]{bai2009panel}
Bai, J. (2009).
\newblock Panel data models with interactive fixed effects.
\newblock {\em Econometrica\/}~{\em 77\/}(4), 1229--1279.

\bibitem[\protect\citeauthoryear{Carvalho, Polson, and Scott}{Carvalho et~al.}{2009}]{carvalho2009handling}
Carvalho, C.~M., N.~G. Polson, and J.~G. Scott (2009).
\newblock Handling sparsity via the horseshoe.
\newblock In {\em Artificial intelligence and statistics}, pp.\  73--80. PMLR.

\bibitem[\protect\citeauthoryear{Duan, Young, Nishimura, and Dunson}{Duan et~al.}{2020}]{duan2020bayesian}
Duan, L.~L., A.~L. Young, A.~Nishimura, and D.~B. Dunson (2020).
\newblock Bayesian constraint relaxation.
\newblock {\em Biometrika\/}~{\em 107\/}(1), 191--204.

\bibitem[\protect\citeauthoryear{Gneiting and Raftery}{Gneiting and Raftery}{2007}]{gneiting2007strictly}
Gneiting, T. and A.~E. Raftery (2007).
\newblock Strictly proper scoring rules, prediction, and estimation.
\newblock {\em Journal of the American statistical Association\/}~{\em 102\/}(477), 359--378.

\bibitem[\protect\citeauthoryear{Jauch, D{\"u}ker, and Hoff}{Jauch et~al.}{2025}]{jauch2025prior}
Jauch, M., M.-C. D{\"u}ker, and P.~Hoff (2025).
\newblock Prior distributions for structured semi-orthogonal matrices.
\newblock {\em arXiv preprint arXiv:2501.10263\/}.

\bibitem[\protect\citeauthoryear{Kimeldorf and Wahba}{Kimeldorf and Wahba}{1970}]{kimeldorf1970correspondence}
Kimeldorf, G.~S. and G.~Wahba (1970).
\newblock A correspondence between {B}ayesian estimation on stochastic processes and smoothing by splines.
\newblock {\em The Annals of Mathematical Statistics\/}~{\em 41\/}(2), 495--502.

\bibitem[\protect\citeauthoryear{Lang and Brezger}{Lang and Brezger}{2004}]{lang2004bayesian}
Lang, S. and A.~Brezger (2004).
\newblock Bayesian p-splines.
\newblock {\em Journal of computational and graphical statistics\/}~{\em 13\/}(1), 183--212.

\bibitem[\protect\citeauthoryear{Lenk and Choi}{Lenk and Choi}{2017}]{lenk2017bayesian}
Lenk, P.~J. and T.~Choi (2017).
\newblock Bayesian analysis of shape-restricted functions using gaussian process priors.
\newblock {\em Statistica Sinica\/}, 43--69.

\bibitem[\protect\citeauthoryear{Lin and Dunson}{Lin and Dunson}{2014}]{lin2014bayesian}
Lin, L. and D.~B. Dunson (2014).
\newblock Bayesian monotone regression using gaussian process projection.
\newblock {\em Biometrika\/}~{\em 101\/}(2), 303--317.

\bibitem[\protect\citeauthoryear{Margaritella, In{\'a}cio, and King}{Margaritella et~al.}{2021}]{margaritella2021parameter}
Margaritella, N., V.~In{\'a}cio, and R.~King (2021).
\newblock Parameter clustering in {B}ayesian functional principal component analysis of neuroscientific data.
\newblock {\em Statistics in Medicine\/}~{\em 40\/}(1), 167--184.

\bibitem[\protect\citeauthoryear{Matuk, Herring, and Dunson}{Matuk et~al.}{2022}]{matuk2022bayesian}
Matuk, J., A.~H. Herring, and D.~B. Dunson (2022).
\newblock Bayesian modeling of nearly mutually orthogonal processes.
\newblock {\em arXiv preprint arXiv:2205.12361\/}.

\bibitem[\protect\citeauthoryear{Nolan, Goldsmith, and Ruppert}{Nolan et~al.}{2025}]{nolan2025bayesian}
Nolan, T.~H., J.~Goldsmith, and D.~Ruppert (2025).
\newblock Bayesian functional principal components analysis via variational message passing with multilevel extensions.
\newblock {\em Bayesian Analysis\/}~{\em 20\/}(1), 1459--1485.

\bibitem[\protect\citeauthoryear{Petralia, Rao, and Dunson}{Petralia et~al.}{2012}]{petralia2012repulsive}
Petralia, F., V.~Rao, and D.~B. Dunson (2012).
\newblock Repulsive mixtures.
\newblock In {\em Proceedings of the 26th International Conference on Neural Information Processing Systems-Volume 2}, pp.\  1889--1897.

\bibitem[\protect\citeauthoryear{Plumlee and Joseph}{Plumlee and Joseph}{2018}]{plumlee2018orthogonal}
Plumlee, M. and V.~R. Joseph (2018).
\newblock Orthogonal gaussian process models.
\newblock {\em Statistica Sinica\/}, 601--619.

\bibitem[\protect\citeauthoryear{Sen, Patra, and Dunson}{Sen et~al.}{2018}]{sen2018constrained}
Sen, D., S.~Patra, and D.~Dunson (2018).
\newblock Constrained inference through posterior projections.
\newblock {\em arXiv preprint arXiv:1812.05741\/}.

\bibitem[\protect\citeauthoryear{Shively, Sager, and Walker}{Shively et~al.}{2009}]{shively2009bayesian}
Shively, T.~S., T.~W. Sager, and S.~G. Walker (2009).
\newblock A bayesian approach to non-parametric monotone function estimation.
\newblock {\em Journal of the Royal Statistical Society Series B: Statistical Methodology\/}~{\em 71\/}(1), 159--175.

\bibitem[\protect\citeauthoryear{Suarez and Ghosal}{Suarez and Ghosal}{2017}]{suarez2017bayesian}
Suarez, A.~J. and S.~Ghosal (2017).
\newblock Bayesian estimation of principal components for functional data.
\newblock {\em Bayesian Analysis\/}~{\em 12\/}(2).

\bibitem[\protect\citeauthoryear{Wakayama and Sugasawa}{Wakayama and Sugasawa}{2024}]{wakayama2024spatiotemporal}
Wakayama, T. and S.~Sugasawa (2024).
\newblock Spatiotemporal factor models for functional data with application to population map forecast.
\newblock {\em Spatial Statistics\/}~{\em 62}, 100849.

\bibitem[\protect\citeauthoryear{Wakayama, Sugasawa, and Kobayashi}{Wakayama et~al.}{2025}]{wakayama2025similarity}
Wakayama, T., S.~Sugasawa, and G.~Kobayashi (2025).
\newblock Similarity-based random partition distribution for clustering functional data.
\newblock {\em Journal of the Royal Statistical Society Series C: Applied Statistics\/}, qlaf037.

\bibitem[\protect\citeauthoryear{Wheeler, Dunson, and Herring}{Wheeler et~al.}{2017}]{wheeler2017bayesian}
Wheeler, M.~W., D.~B. Dunson, and A.~H. Herring (2017).
\newblock Bayesian local extremum splines.
\newblock {\em Biometrika\/}~{\em 104\/}(4), 939--952.

\bibitem[\protect\citeauthoryear{Williams and Rasmussen}{Williams and Rasmussen}{2006}]{williams2006gaussian}
Williams, C.~K. and C.~E. Rasmussen (2006).
\newblock {\em Gaussian processes for machine learning}, Volume~2.
\newblock MIT press Cambridge, MA.

\bibitem[\protect\citeauthoryear{Xie and Xu}{Xie and Xu}{2020}]{xie2020bayesian}
Xie, F. and Y.~Xu (2020).
\newblock Bayesian repulsive {G}aussian mixture model.
\newblock {\em Journal of the American Statistical Association\/}~{\em 115\/}(529), 187--203.

\bibitem[\protect\citeauthoryear{Yao, M{\"u}ller, and Wang}{Yao et~al.}{2005}]{yao2005functional}
Yao, F., H.-G. M{\"u}ller, and J.-L. Wang (2005).
\newblock Functional data analysis for sparse longitudinal data.
\newblock {\em Journal of the American statistical association\/}~{\em 100\/}(470), 577--590.

\bibitem[\protect\citeauthoryear{Yu, Li, Noe, Fischer-Baum, and Vannucci}{Yu et~al.}{2023}]{yu2023bayesian}
Yu, C.-H., M.~Li, C.~Noe, S.~Fischer-Baum, and M.~Vannucci (2023).
\newblock Bayesian inference for stationary points in gaussian process regression models for event-related potentials analysis.
\newblock {\em Biometrics\/}~{\em 79\/}(2), 629--641.

\bibitem[\protect\citeauthoryear{Yu, Suresh, Choromanski, Holtmann-Rice, and Kumar}{Yu et~al.}{2016}]{yu2016orthogonal}
Yu, F. X.~X., A.~T. Suresh, K.~M. Choromanski, D.~N. Holtmann-Rice, and S.~Kumar (2016).
\newblock Orthogonal random features.
\newblock {\em Advances in neural information processing systems\/}~{\em 29}.

\bibitem[\protect\citeauthoryear{Zhang, Cao, and Wang}{Zhang et~al.}{2025}]{zhang2025robust}
Zhang, J., J.~Cao, and L.~Wang (2025).
\newblock Robust {B}ayesian functional principal component analysis.
\newblock {\em Statistics and Computing\/}~{\em 35\/}(2), 46.

\end{thebibliography}

\end{document}